\documentclass[twocolumn,aps,prd,floats]{revtex}
\usepackage{graphics}
\IfFileExists{url.sty}{\usepackage{url}}
                      {\newcommand{\url}{\texttt}}

\makeatletter

\providecommand{\LyX}{L\kern-.1667em\lower.25em\hbox{Y}\kern-.125emX\@}
\let\SF@@footnote\footnote
\def\footnote{\ifx\protect\@typeset@protect
    \expandafter\SF@@footnote
  \else
    \expandafter\SF@gobble@opt
  \fi
}
\expandafter\def\csname SF@gobble@opt \endcsname{\@ifnextchar[
  \SF@gobble@twobracket
  \@gobble
}
\edef\SF@gobble@opt{\noexpand\protect
  \expandafter\noexpand\csname SF@gobble@opt \endcsname}
\def\SF@gobble@twobracket[#1]#2{}

\usepackage[T1]{fontenc}
\usepackage{graphics}

\makeatletter

\newcommand{\cmb}{cosmic microwave background }
\newcommand{\dx}{\partial x}
\def\dy{\partial y}

\def\vgam{\vec{\alpha}}
\def\valpha{\vec{\alpha}}
\def\vxi{\vec{\xi}}

\def\Bobs{\hat{B}}
\def\Qobs{\hat{Q}}
\def\Uobs{\hat{U}}

\def\vl{\vec{l}}

\draft

\makeatother

\makeatother

\begin{document}

\twocolumn[\hsize\textwidth\columnwidth\hsize\csname @twocolumnfalse\endcsname

\title{Cosmic string lens effects on CMB polarization patterns}

\author{K. Benabed and F. Bernardeau}

\address{Service de Physique Th\'{e}orique, C.E. de Saclay, 91191 Gif-Sur-Yvette, France.}

\date{\today{}}

\maketitle
\widetext

\begin{abstract}
Extending the Kaiser-Stebbins mechanism we propose here a method for detecting
relics of topological defects such as cosmic strings based on lens-induced small-scale
\( B \)-type polarization in the \cmb (CMB). Models of inflation, in which
large scale structures of the Universe emerge from the inflaton fluctuations,
do not exclude the formation of topological defects at the end of the inflationary
phase. In such case, we show that the lens effect of a string on the small scale
\( E \)-type polarization of the \cmb induces a significant amount of \( B \)-type
polarization along the line-of-sight. The amplitude of the effect is estimated
for different resolutions of \cmb experiments. 
\end{abstract}
\vspace{0.2cm}

] \narrowtext

\section*{Introduction}

The temperature anisotropies of the \cmb offer a unique window towards the physics
of the early Universe and for the understanding of the large-scale structures.
Current observations of the temperature anisotropies power spectrum, \( C_{\ell } \),
point toward the existence of a well localized first acoustic peak\cite{firstacousticpeak}.
If this result is confirmed by the next generation of CMB experiments, it supports
models of large-scale structure formation from adiabatic scalar fluctuations
at the expense of models of topological defects and more particularly of cosmic
strings \cite{Kibble}\cite{stringCl}. Furthermore, the shape and normalization
of the local matter density power spectrum, \( P(k) \), is also in bad agreement
with the CMB data for such models \cite{Albrecht}. This suggests that \emph{only}
a small fraction of the large-scale inhomogeneities might be due to topological
defects. However, recent studies have shown that in realistic models of inflation
cosmic string formation seems quite natural at the end of the inflationary period:
it is a natural outcome in Super-Symmetry inspired scenario\cite{Deffayet};
it can also be obtained during a pre- or reheating process \cite{Preheat}. 

The effects of cosmic strings on the last scattering surface temperature map
have been described by Kaiser \& Stebbins \cite{KSeffect}. If a cosmic string
is moving against an homogeneous surface of uniform temperature, the energy
of the deflected photon \cite{Vilenkin} is enhanced or reduced (the photons
are then blueshifted or redshifted) depending on whether the photon is passing
behind or ahead the moving string, a mechanism through which temperature anisotropies
are generated. The aim of this paper is to explore the possibilities of having
similar effects for the polarization properties. Obviously, if the background
surface is unpolarized, the deflected photons remain unpolarized and no effects
can be observed. However if the background sky is polarized then the polarization
pattern is affected through lens effects and, in particular a geometrical deformation
can naturally induce \( B \)-type polarization out of the \( E \) component.
This mechanism has been described for the large-scale structures \cite{LSS-BPol}
and recognized as a major source of \( B \)-type small-scale CMB polarization.
We are interested here in the case of cosmic strings for which this effect can
be easily investigated and visualized. Note that, by doing so, we neglect a
possible coupling with an axionic field associated with the string that could
induce significant photon-string non-gravitational couplings to a finite distance\cite{HarveyNaculich}.
The background model of large-scale structure formation of this paper is therefore
inflation driven adiabatic fluctuations with a few cosmic strings that may have
survived from a late time phase transition although with a significant linear
energy density. Note that what we are describing here is a secondary effect
from a perturbation theory point of view in the sense that it is quadratic in
the metric perturbation: it is a coupling between the local gravitational potential
and the potential on the last-scattering surface (the Kaiser-Stebbins effect
is also a secondary effect). We do not attempt to describe the primary anisotropies
induced at the recombination time that have been examined in other studies\cite{StringCMBPol}.

In sect. 1 we examine in detail the effects induced by straight strings and
by circular loops. Sect. 2 contains the result of simulations of the effect.
Then, in sect. 3, we estimate the detectable amplitude of the \( B \) component
in the case of a straight string driven deformation.

\vspace{0.4cm}

\section{Cosmic string lens effect}

In inflationary scenario, at any given scale, scalar perturbations give birth
to a scalar polarization pattern -- that is to say, to \( E \)-type polarization
-- whereas tensor modes, that can induce both \( E \) and \( B \)-type polarization,
contribute only at very large scale \cite{noBpol}. This result accounts for
the symmetry of the fluctuations. It implies that at small scales, the pseudo-scalar
\( B \) field,

\begin{equation}
\label{eqB}
B\equiv \Delta ^{-1}[(\dx ^{2}-\dy ^{2})\, U-2\dx \dy \, Q],
\end{equation}
 defined\footnote{%
Throughout the paper we work in the small angular scale limit. See \cite{AllSkyPolarization}
for a general discussion of these properties. 
} from the Stokes parameters \( Q \) and \( U \) is zero. The polarization
field is then entirely defined by the scalar \( E \) field,

\begin{equation}
E\equiv \Delta ^{-1}[(\dx ^{2}-\dy ^{2})\, Q+2\dx \dy \, U].
\end{equation}

Since the polarization vector is parallel transported along the geodesics, a
gravitational lens affects the polarization simply by displacing the apparent
position of the polarized light source\cite{Schneider}. In other words, the
observed Stokes parameters \( \Qobs  \) and \( \Uobs  \) are given in terms
of the \emph{primary} (i.e. unlensed) ones by:

\begin{equation}
\label{eqDep}
\Qobs (\vgam )=Q(\vgam +\vxi ),\qquad \Uobs (\vgam )=U(\vgam +\vxi ).
\end{equation}
 where \( \vxi  \) is the displacement field at angular position \( \valpha  \)
(\( \valpha  \) is a 2D vector that gives the sky coordinates in the small
angle limit). The displacement, \( \vxi  \), is given by the integration of
the gravitational potential along the line-of-sights. We will assume in this
paper that the only potential acting as lens is the cosmic string potential.
It obviously depends on the shape, equation of state and dynamics of the string.

Putting (\ref{eqDep}) in (\ref{eqB}), we can write a general expression of
\( \Delta \Bobs  \) in presence of lenses;

\begin{equation}
\begin{array}{l}
\Delta \Bobs (\vgam )=-2Q_{,ij}(\vgam +\vxi )(\delta ^{i}_{x}+\xi _{,x}^{i})(\delta _{y}^{j}+\xi ^{j}_{,y})\\
-2Q_{,i}(\vgam +\vxi )\xi _{,xy}^{i}\\
+U_{,ij}(\vgam +\vxi )\\
\: \: \: \times [(\delta _{x}^{i}+\xi _{,x}^{i})(\delta _{x}^{j}+\xi _{,x}^{j})-(\delta _{y}^{i}+\xi _{,y}^{i})(\delta _{y}^{j}+\xi _{,y}^{j})]\\
+U_{,i}(\vgam +\vxi )(\xi _{,xx}^{i}-\xi ^{i}_{,yy}).
\end{array}
\end{equation}
 There are no reasons for the displacement field to preserve a non-zero \( B \)-type
\cmb polarization simply because the two scalar field composition (one being
the primary scalar perturbation, the other the line-of-sight gravitational potential)
breaks the parity invariance. 

For illustration we examine here explicitl{\small y} the two special cases of
a straight cosmic string and of a circular cosmic string, both of them in a
plane orthogonal to the line-of-sight.

\subsection{The case of a straight string}

Let us assume that a straight string is aligned along the \( y \)-axis. Then
the displacement is uniform at each side of the string. The deflection angle,
\( \alpha =4\pi G\mu  \) \cite{Vilenkin} (where \( G \) is the Newton constant
and \( \mu  \) the string linear energy density) induces a displacement given
by, 
\begin{equation}
\xi _{x}=\pm \xi _{0},\quad \xi _{0}=4\pi G\mu \frac{{\mathcal{D}}_{\textrm{CMB},\textrm{string}}}{{\mathcal{D}}_{\textrm{string}}},
\end{equation}
 the sign depending on which side of the string one observes; the displacement
along \( y \) is obviously 0. \( {\mathcal{D}}_{\textrm{CMB},\textrm{string}} \)
and \( {\mathcal{D}}_{\textrm{string}} \) are the cosmological angular distances
between, respectively, the last scattering surface and the string, and the string
and the observer. In the following, we will assume that we are in the most favorable
case for detection, when the ratio of the distance is about unity, hence removing
any geometrical dependence on the cosmological parameters. Then, the string
lays at equal distance between the last scattering surface and the observer\footnote{%
this means a redshift of 3 for an Einstein-de Sitter universe. 
}. For a \( G\mu  \) around \( 10^{-6} \) \cite{Vilenkin}, the typical expected
displacement is about less than 10 arc seconds.

We can write the expression of the Stokes parameters, 
\begin{eqnarray}
\Qobs (x,y) & = & Q(x-\xi _{0},y)\, \theta (x-x_{0})\nonumber \\
 & + & Q(x+\xi _{0},y)\, \left( 1-\theta (x-x_{0})\right) ,
\end{eqnarray}
 where \( \theta  \) is the step function, \( x_{0} \) is the position of
the string. The same expression holds for \( \Uobs  \). Since the primary polarization
map is \( B \) free, the Laplacian of the observable \( B \) field is finally
given by, 
\begin{eqnarray}
\Delta \Bobs (\vgam ) & = & \delta (x-x_{0})\left( \left| U_{,x}\right| _{\! -}^{\! +}-2\left| Q_{,y}\right| _{\! -}^{\! +}\right) \nonumber \\
 & + & \delta '(x-x_{0})\left| U\right| _{\! -}^{\! +},\label{Bobsstraight} 
\end{eqnarray}
 where we define 
\begin{equation}
\left| X\right| _{\! -}^{\! +}\equiv \hat{X}(x_{0}^{+})-\hat{X}(x_{0}^{-})=X(x_{0}-\xi _{0})-X(x_{0}+\xi _{0}).
\end{equation}
 One can note in Eq. (\ref{Bobsstraight}) that the effect is entirely due to
the discontinuity induced by the string on the polarization map. Furthermore,
the \( B \) component of the polarization will only be non-zero on the string
itself. Obviously, the efficiency with which such an effect will be observed
depends on the angular precision of the detectors as we discuss later.

\subsection{The case of a circular string}

The case of a collapsing circular string loop allows to complement our analysis
with the effect of the string curvature. As shown in \cite{deLaix} the lens
effect of such a string, when facing the observer, is equivalent to the one
of a static linear mass distribution. The structure of the displacement field
is then simple. Let us consider a loop centered at the origin of our coordinate
and of radius \( \alpha _{l} \). If one observes towards a direction through
the loop, the displacement is nil. Outside the loop, the displacement decreases
as \( {\alpha _{l}}/{\alpha } \). We have 
\begin{equation}
\vxi (\vgam )=-2\xi _{0}\frac{\alpha _{l}}{\alpha ^{2}}\vgam \: \, \, \, \, \mathrm{for}\: \, \, \, \, \alpha >\alpha _{l}.
\end{equation}
 Note that in \( \xi _{0} \), \( \mu  \) is an effective quantity that contains
the effects of dynamics as well. Then, \( \Qobs  \) is 
\begin{eqnarray}
\Qobs (\vgam ) & = & Q\left[ \vgam \left( 1-2\xi _{0}\frac{\alpha _{l}}{\alpha ^{2}}\right) \right] \, \theta (\alpha -\alpha _{l})\nonumber \\
 & + & Q(\vgam )\, \left[ 1-\theta (\alpha -\alpha _{l})\right] .
\end{eqnarray}
 Two effects are induced in this case. The first one, also present in the straight
string case, comes from the discontinuity of the polarization field; this is
the \emph{strong lensing} effect. It is due to the existence of a critical region
in the source plane where objects can have multiple images (two in this case,
but it can be more in general \cite{deLaix}). The second one is a \emph{weak
lensing} effect simply due the deformation of the source plane; it will be small
compared to the other. This latter effect is investigated in more detail in
\cite{CestMoi}. We expect these two effects to be present for any string model.

\section{Simulated maps}

We present in Figs. \ref{FiveMinB}-\ref{OneMinE} the simulation results of
circular cosmic string (30 arc minutes radius, \( \xi _{0}=5'' \)). The \cmb polarization
realization uses \( C_{\ell } \) calculated with a standard \( \Lambda \mathrm{CDM} \)
model. Only scalar primary perturbations are used here since we do not expect
any significant tensor mode at such a small scale; without the string, there
is no signal in the \( B \) component.  

The hot and cold (black and white) patches run along the string path. They come
from the \( \delta ' \) term in the eq. (\ref{Bobsstraight}). Its amplitude
is the result of a finite difference in the \( U \) field at distance \( 2\xi _{0} \)\emph{.}
We will see in last section that, at small filtering resolution, this term dominates
the amplitude of \( B \) polarization.
\begin{figure}
{\par\centering \resizebox*{8cm}{8cm}{\includegraphics{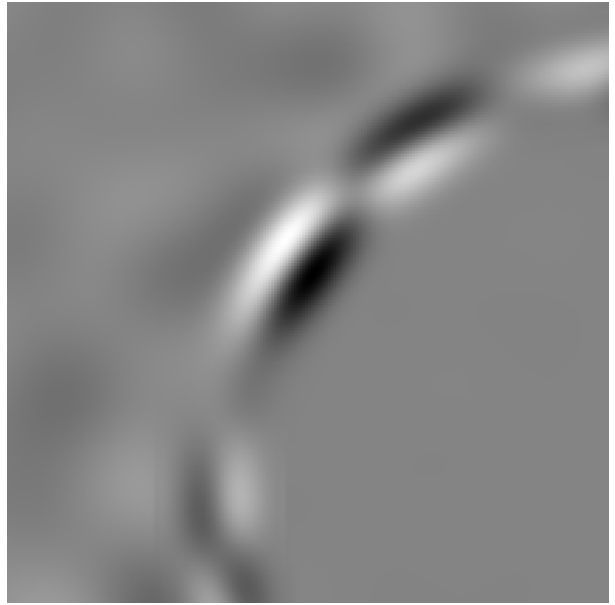}} \par}

\vspace{.3cm}

\caption{\label{FiveMinB} \protect\( B\protect \) field for a circular loop crossing
a \protect\( 50'\times 50'\protect \) window. The filter resolution is 5 arc
minutes. At this scale, the \protect\( B\protect \) field is less than \protect\( 1\%\protect \)
the \protect\( E\protect \) one. The very faint patches that can be noticed
above the string are \emph{weak lensing} effect signature (a few percent of
the \emph{strong lensing} effect coming from the critical region). }
\end{figure}
 
\begin{figure}
{\par\centering \resizebox*{8cm}{8cm}{\includegraphics{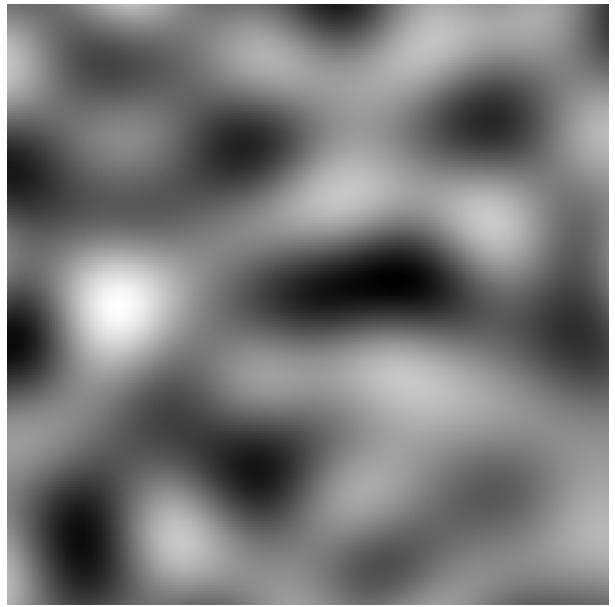}} \par}

\vspace{.3cm}

\caption{\label{FiveMinE} Same as Fig. \ref{FiveMinB} for the \protect\( E\protect \)
field. At this scale the string remains completely ``diluted'' in the \protect\( E\protect \)
field and cannot be seen (the effect is smaller than 1\% of the mean primary
\protect\( E\protect \) signal). }
\end{figure}

It is interesting to note here that the \emph{weak lensing} effect is negligible
at these scales. Besides, even the discontinuity effect is really small at a
5 arc minute angular scale. A 4 times better resolution enhances significantly
the signal (about 10 times). Note also that the hot and cold spots along the
strings have the same linear sizes as the typical peaks of the polarization
field. We expect that this feature and the very peculiar shape of the effect
on the \cmb polarization maps could help discriminating between this effect
and other secondary polarization sources or foregrounds (lensing from large-scale
structures, dust polarization...) The extraction of a clean \cmb polarization out
of a signal with foregrounds has been studied\cite{prunetbouchet} at larger
scale. Few is known about contamination of the \( B \) signal at the scale
we are looking at here.
\begin{figure}
{\par\centering \resizebox*{8cm}{8cm}{\includegraphics{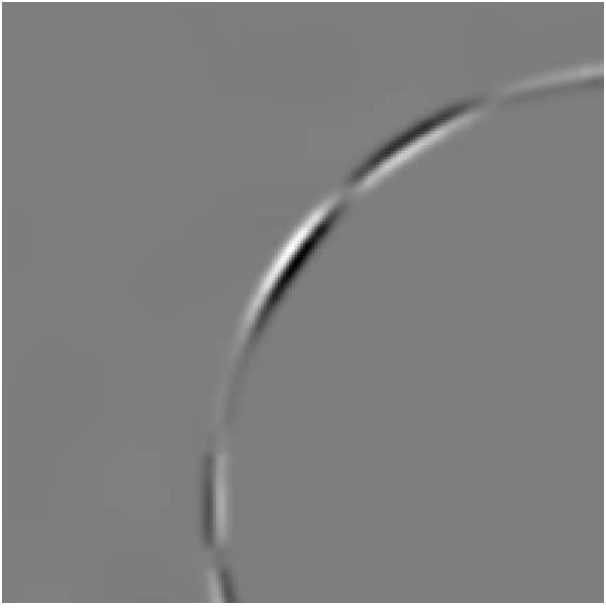}} \par}

\vspace{.3cm}

\caption{\label{OneMinB} Same as \ref{FiveMinB} with a better resolution (1.2'). The
discontinuity effect is less diluted. The \protect\( B\protect \) field is
now in amplitude about 10\% of the typical \protect\( E\protect \) fluctuations. }
\end{figure}
 
\begin{figure}
{\par\centering \resizebox*{8cm}{8cm}{\includegraphics{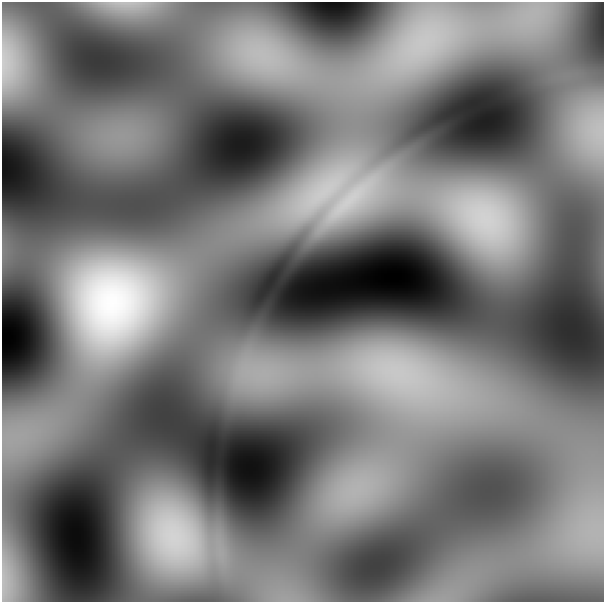}} \par}

\vspace{.3cm}

\caption{\label{OneMinE} Same as \ref{FiveMinE} for the 1.2' resolution. The discontinuity
effect is now visible in \protect\( E\protect \). Lens effects induce mode
couplings that create structures at very small scales dominating over primary
structures.}
\end{figure}

\section{Amplitude of a Straight String effect }

We come back to the case of a straight cosmic string. It is easy to estimate
the amplitude of the effect which only consists of a discontinuity. At small
angles we can decompose the \( E \) field in plane wave Fourier modes, 
\begin{equation}
E(\vgam )=\int \frac{{\mathrm{d}}^{2}l}{2\pi }\tilde{E}(\vl ){\mathrm{e}}^{{\mathrm{i}}\vgam .\vl }.
\end{equation}
 It is straightforward to write for \textbf{\( B \)}, in Fourier space, 
\begin{eqnarray}
\Delta \Bobs (x,y) & = & 2\int \frac{{\mathrm{d}}^{2}l}{2\pi }\tilde{E}(\vl ){\mathrm{e}}^{{\mathrm{i}}\vgam .\vl }\left( {\mathrm{e}}^{{\mathrm{i}}\xi _{0}l_{x}}-{\mathrm{e}}^{-{\mathrm{i}}\xi _{0}l_{x}}\right) \nonumber \\
 &  & \! \! \! \! \! \! \! \! \! \times \left[ \frac{l_{x}l_{y}}{l^{2}}\delta '(x-x_{0})+{\mathrm{i}}\frac{l_{y}^{3}}{l^{2}}\delta (x-x_{0})\right] .
\end{eqnarray}
 This expression makes sense only if convolved with a \emph{test function};
that is to say convolved with a suitable window function. For simplicity, we
assume that our observational device window is described by a Gaussian window
function \( W \) of width \( \alpha _{w} \), 
\begin{equation}
W(\vgam )=\frac{1}{2\pi \alpha _{w}}{\mathrm{e}}^{-\frac{\alpha ^{2}}{2\alpha _{w}^{2}}},\: \tilde{W}(k_{x},k_{y})={\mathrm{e}}^{-\frac{\alpha _{w}^{2}\, (k_{x}^{2}+k_{y}^{2})}{2}}.
\end{equation}
 Then we have, 
\begin{eqnarray}
\Delta \Bobs _{W}(x,y) & = & 2\int \frac{{\mathrm{d}}^{2}l}{2\pi }\frac{{\mathrm{d}}k}{2\pi }\tilde{E}(\vl ){\mathrm{e}}^{{\mathrm{i}}\left[ x_{0}(l_{x}-k)+xk+yl_{y}\right] }\nonumber \\
 &  & \! \! \! \! \! \! \! \! \! \! \! \! \! \! \! \! \! \! \! \! \! \! \! \! \! \! \! \! \times \tilde{W}(k,l_{y})\left( {\mathrm{e}}^{{\mathrm{i}}\xi _{o}l_{x}}-{\mathrm{e}}^{-{\mathrm{i}}\xi _{o}l_{x}}\right) \left[ {\mathrm{i}}\frac{l_{x}l_{y}k}{l^{2}}+{\mathrm{i}}\frac{l_{y}^{3}}{l^{2}}\right] .
\end{eqnarray}
 The r.m.s. of \( \Delta \Bobs _{W} \) can then be expressed as a function
of the \( E \) power spectrum \( C_{E}(l) \), 
\begin{eqnarray}
\left\langle \left( \Delta \Bobs _{W}\right) ^{2}\right\rangle  & = & 2{\mathrm{e}}^{-\frac{(x-x_{o})^{2}}{\alpha _{w}^{2}}}\int \frac{{\mathrm{d}}^{2}l}{\pi ^{3}}C_{E}(l)\sin ^{2}(\xi _{0}l_{x})\nonumber \label{test} \\
 &  & \! \! \! \! \! \! \! \! \! \! \! \! \! \times {\mathrm{e}}^{-l_{y}^{2}\alpha _{w}^{2}}\left[ \frac{l_{x}^{2}l_{y}^{2}(x-x_{0})^{2}}{l^{4}\alpha _{w}^{6}}+\frac{l_{y}^{6}}{l^{4}\alpha _{w}^{2}}\right] .\label{DBobsW} 
\end{eqnarray}
 \( C_{E}(l), \) has a natural cutoff due to the Silk damping scale, \( l_{\textrm{damp }} \)(\( 1/l_{{\mathrm{damp}}}\sim 10' \)),
a scale much bigger than the induced displacement. Therefore, we can replace
\( \sin ^{2}(\xi _{0}l_{x}) \) by its expansion \( \xi _{0}^{2}l_{x}^{2} \).
Then the amplitude of the effect grows like \( \xi _{0} \). Besides, if the
size of the window is smaller than the typical scale of \( \Delta E \) structures,
\( \alpha _{w}\ll \alpha _{\textrm{peaks}} \)with \( \alpha _{\textrm{peaks}}\sim 10^{-3} \),
we have \( \exp (-l_{y}^{2}\alpha _{w}^{2})\sim 1 \). And from Eq. (\ref{DBobsW})
we can then calculate: 
\begin{eqnarray}
\begin{array}{l}
\frac{\left\langle \left( \Delta \Bobs _{w}(x=x_{0}\pm \alpha _{w},y)\right) ^{2}\right\rangle ^{1/2}}{\left\langle \left( \Delta E\right) ^{2}\right\rangle ^{1/2}}=\frac{1}{4\sqrt{\pi \mathrm{e}}}\frac{\xi _{0}}{\alpha _{w}}\sqrt{5+8\frac{\alpha ^{2}_{\textrm{peaks}}}{\alpha ^{2}_{w}}}.
\end{array}\label{eqapprox} 
\end{eqnarray}
 The distance to the string, \( x=x_{0}\pm \alpha _{w} \), have been chosen
to give a realistic account of the effect (in the simulation this correspond
almost to the peak of the hot and cold patches). Fig.\ref{approx} shows these
results for a \( \Lambda \mathrm{CDM} \) model -- the only dependence with
the cosmological parameters appears in the position of the polarization peaks
which depends essentially on the global curvature of the Universe. Our approximation
is not exact at 5', but is fairly accurate at 1'. Numerically, we found that,
at 5' resolution, the amplitude of the effect evolves like \( \sim 325\, \xi _{0} \)
at the position \( x=x_{0}\pm \alpha _{w} \); it is \( \sim 1.5\times 10^{-2} \)
when \( \xi _{0}=10'' \). The slope at small \( \alpha _{w} \) is due to the
\( \alpha ^{2}_{\textrm{peaks}}/\alpha ^{2}_{w} \) in eq. (\ref{eqapprox})
that is to say to the \( \delta ' \) in eq. (\ref{Bobsstraight}). This is
in good agreement with the simulations that suggested an effect of this order
of magnitude dominated, at small \( \alpha _{w} \), by the finite difference
in \( U \) field at \( 2\xi _{0} \) scale. 
\begin{figure}
{\par\centering \resizebox*{8cm}{!}{\includegraphics{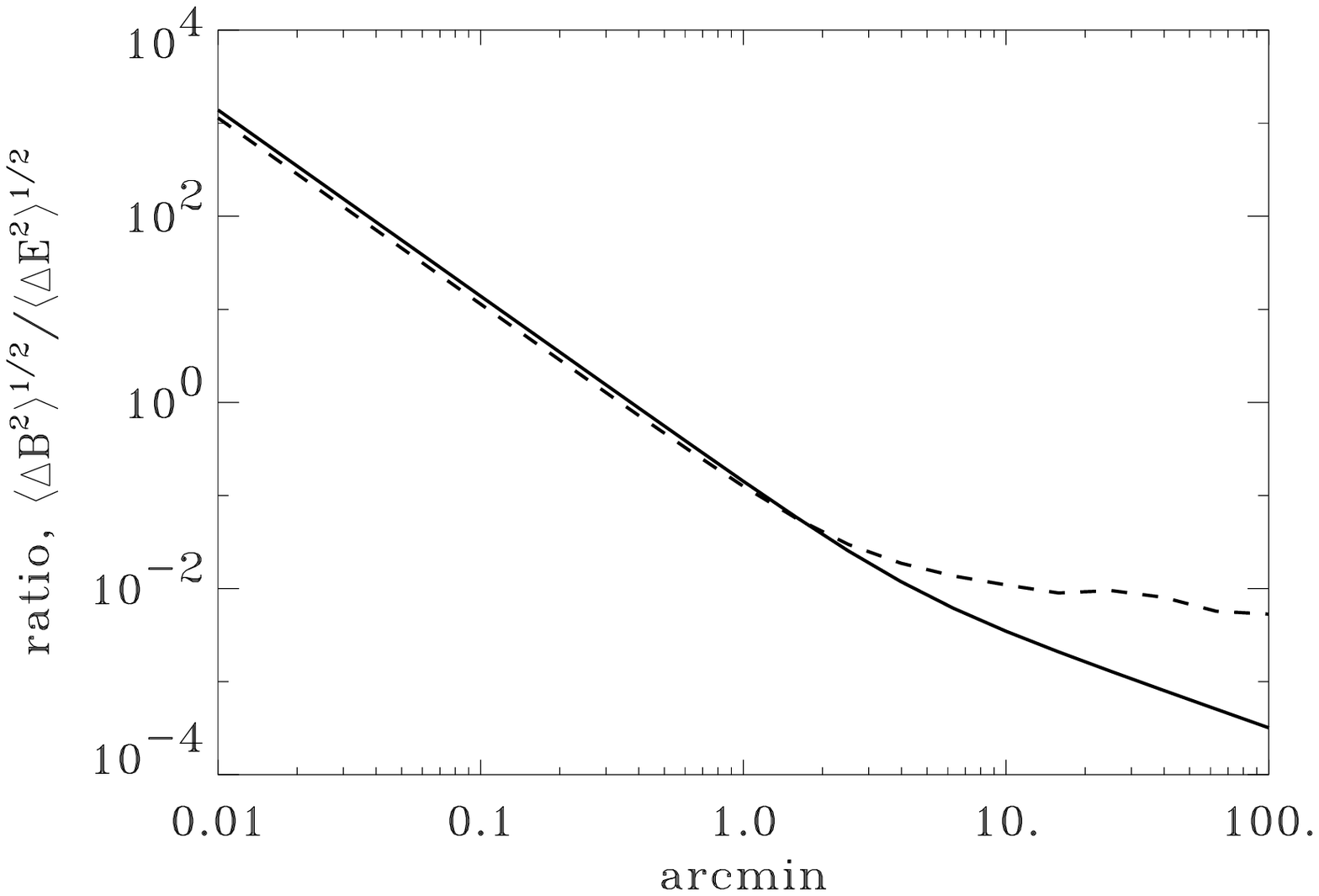}} \par}

\caption{\label{approx} A comparison of the exact computation of \protect{}}

\( \sqrt{\langle \Delta B^{2}\rangle /\langle \Delta E^{2}\rangle }\protect  \)
(dashed line) at \protect\( x=x_{0}\pm \alpha _{w}\protect  \) and \protect\( \xi _{0}=10''\protect  \)
and its approximation (see eq. \ref{eqapprox}). The amplitude of the effect
is about 10\% at 1' scale, about 1.5\% at 5'. The agreement between the exact
amplitude and eq. (\ref{eqapprox}) weakens for scales above \protect\( 2\sim3' \protect  \). 
\end{figure}

\section*{Conclusions}

The so far the planned \cmb experiments will have, at their very best, a 5'
resolution \cite{Plancketal}. We showed that at this angular scale and assuming
that it exists a string with \( \xi _{0}=10'' \) (which is perhaps somewhat
enthusiastic), we can expect a signal in \( B \)-type polarization with an
amplitude of about 1\% the signal in \( E \)-type polarization. This signal
is too weak to be actually detected. Beside, the weak lensing is expected to
induce \( B \)-type polarization at the same scale and is likely to hide any
string effect. Improving the resolution of the detector however will dramatically
enhance the detectability of a string effect; at 1' scale, we expect indeed
to gain a factor 10 in the amplitude of the effect that should make the detection
possible. The detection of cosmic strings through this effect (or through the
Kaiser-Stebbins effect which also requires a good angular resolution) will probably
be possible only with the post Planck generation of instruments\cite{PostPlanck}.

\acknowledgements

The authors would like to thank P. Peter, L. Kofman and especially J.P. Uzan
for encouraging discussions and comments on the manuscript. We also thank A.
Riazuelo for the use of his Boltzmann CMB code.

\end{document}